\documentclass[journal,comsoc]{IEEEtran}

\usepackage[T1]{fontenc}
\usepackage{soul}
\usepackage{hyperref}
\usepackage{tikz}
\usetikzlibrary{arrows.meta}
\usepackage[cmintegrals]{newtxmath}
\usepackage{graphicx}
\usepackage[utf8]{inputenc}
\usepackage{csquotes}
\usepackage[font=footnotesize,labelfont=bf]{caption}
\usepackage{adjustbox}
\renewenvironment{IEEEbiography}[1]
  {\IEEEbiographynophoto{#1}}
  {\endIEEEbiographynophoto}
\usepackage{enumitem}
\makeatletter
\renewcommand{\fnum@figure}{Figure \thefigure}
\makeatother
\begin{document}

\title{Trustworthy Digital Twins in the Industrial Internet of Things with Blockchain}

\author{Sabah Suhail, Rasheed Hussain, Raja Jurdak, and Choong Seon Hong
\thanks{S. Suhail and C. S. Hong are with Department of Computer Science and Engineering, 
Kyung Hee University, South Korea;
(e-mail: (sabah, cshong)@khu.ac.kr).}
\thanks{R. Hussain is with Networks and Blockchain Lab,
Innopolis University, Russia;
(e-mail: r.hussain@innopolis.ru).}
\thanks{R. Jurdak is with Trusted Networks Lab, School of Computer Science, Queensland University of Technology, Australia;
(e-mail: r.jurdak@qut.edu.au).}
}

\maketitle

\begin{abstract}
Industrial processes rely on sensory data for critical decision-making processes. Extracting actionable insights from the collected data calls for an infrastructure that can ensure the \textit{trustworthiness} of data. To this end, we envision a blockchain-based framework for the Industrial Internet of Things (IIoT) to address the issues of data management and security. Once the data collected from trustworthy sources are recorded in the blockchain,
product lifecycle events can be fed into data-driven systems for process monitoring, diagnostics, and optimized control. In this regard, we leverage Digital Twins (DTs) that can draw intelligent conclusions from data by identifying the faults and recommending precautionary measures ahead of critical events. Furthermore, we discuss the integration of DTs and blockchain to target key challenges of disparate data repositories, untrustworthy data dissemination, and fault diagnosis. Finally, we identify outstanding challenges faced by the IIoT and future research directions while leveraging blockchain and DTs.
\end{abstract}

\IEEEpeerreviewmaketitle

\section{Introduction}\label{introduction}
The Industrial Internet of Things (IIoT) allows the interconnection of the physical and digital worlds resulting in the realization of Cyber-Physical Systems (CPS) deployed at manufacturing industries. Starting from the transformation of raw materials to end products, the product lifecycle comprises several complex phases, for instance, on-going intricate industrial processes and supply chain trade events. Due to the involvement of multiple entangled entities, industries have to consider several factors for maximizing their revenues, such as cost management, risk prognosis, and quality assurance~\cite{suhail2019orchestrating}. Currently, a plethora of enabling technologies are revolutionizing the industry, for instance, IoT, 3D simulations, High-Performance Computing (HPC), and Distributed Ledger Technology (DLT)~\cite{yaqoob2020blockchain}. However, the overarching question in industrial infrastructure is: \emph{How to unlock the full potential of the enabling technologies to maximize industrial efficiency?} One of the promising solutions is to create a digital fingerprint (or virtual replica) of the underlying product, process, or service where all operations must be analyzed, predicted, and optimized before its real-world implementation. Following a closed-loop, the simulation data are fed back to the physical system to calibrate the operations and to enhance the system performance. Such a bi-directional mapping between \emph{physical space} and \emph{virtual space} is called \emph{Digital Twin} (DT)~\cite{tao2018digital}. As physical entities start to work, DTs integrate data from multiple sources, such as sensor data, historical production data (acquired from product lifecycle data), and domain knowledge to generate comprehensive data (in the form of models or simulations).
During industrial operations, DTs run synchronously with physical counterparts where the primary objective is to track \emph{data inconsistencies} between physical and virtual space. Inconsistencies between the two spaces call for the adoption of better calibration and testing strategies that evolve DT models and physical counterparts in order to support more accurate estimation, optimization, and prediction of the industrial processes~\cite{tao2017digital}.

Data consistency between continuous physical space and discrete virtual space is imperative for DTs since it helps in eliminating operational disruptions and uncertainties. In other words, the data fed into the DTs is of critical significance as it accounts for rational decision making and precise execution. Therefore, the key question is: \emph{How to ensure the trustworthiness of data collated from disparate data silos or participating entities?} In this regard, the nascent technology of blockchain gained ground to address challenging issues in the industry pertaining to product lifecycle data management and data security. Leveraging blockchain allows industries to manage data on a distributed ledger that is shared among all participating entities, not owned by anyone, solve important glitches in traceability, and record events in a secure, immutable, and irrevocable way~\cite{deloitte}.

\begin{figure*}[ht!]
\centerline{\includegraphics[width=6.5in]{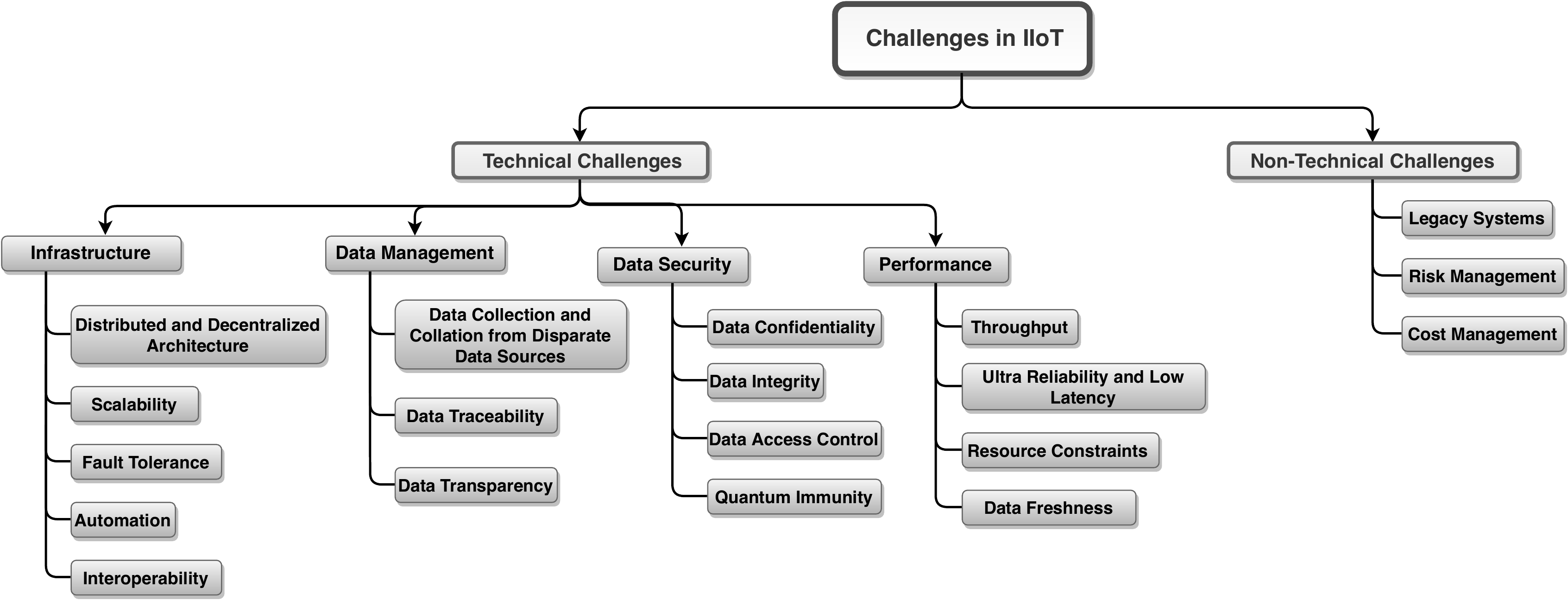}}
\caption{Technical and non-technical challenges in IIoT.}
\label{fig:challenges}
\end{figure*}

In this article, we propose a framework to address the challenges of data management, data security, and predictive maintenance in IIoT. Our main contributions are as follows:
\begin{itemize}
\item To ensure the trustworthiness of data as well as data-generating sources, we use data provenance and data twinning.
\item Considering the challenging requirements of real-time industrial processes, the growing number of participating entities, and threats to the public-key cryptosystem, we focus on the adoption of a lightweight, scalable, and quantum-immune blockchain-based solution for IIoT. 
\item Based on the closed-loop operation between the digital-physical mapping, we enforce calibration and scheduling services at virtual space and physical space respectively. 
\item Finally, we highlight the research challenges that impede the adoption of blockchain and DTs in the IIoT. 
\end{itemize}

\section{Challenging Requirements of IIoT} 
\label{design}
While designing a twinned system for IIoT, we encounter several intriguing questions: (1) how to trust the data generating sources in addition to the data flowing between the physical space and the virtual space? (2) how to effectively manage data storage? and (3) what will be the consequences of conflicts between high concurrency and low throughput? Envisaging these primitive concerns, in this section, we highlight several technical and non-technical challenges (summarized in Fig.~\ref{fig:challenges}) that must be evaluated before designing an effective solution for the twinned data in IIoT. 

\subsection{Technical Challenges}
Among \textit{infrastructure-related challenges} in IIoT, the first challenge is to avoid a single point of failure in centralized architectures and to provide a distributed and decentralized infrastructure for geographically dispersed industrial units. Considering the collaboration among multiple participating entities in IIoT, the second challenge is to organize the industrial system, keeping in view the expected increase in the number of actors (sensors, actuators, equipment, human resources) and activities (trade events, work in process) in the network over time without running into scalability issues. Such complex industrial systems have to consider the on-going industrial processes at different manufacturing units while procuring product components or parts from various competitive vendors. The real-time industrial processes are subjected to stochastic disturbances due to the complexity and dynamism of underlying (sub)-systems, hidden vulnerabilities, machinery malfunctioning, human errors, knowledge gap, missing or erroneous data, etc. Either time-sensitive or latency-tolerant, such manufacturing disturbances require due attention to resume the process. Therefore, the third challenge is the deployment of resilient IoT solutions that can tolerate and recover from accidental or malicious failures without causing data or service unavailability. The fourth challenge is to deploy automation technologies to speed up and streamline underlying processes in IIoT. For instance, the deployment of sensors devices to monitor on-going events, robots, and drones to perform repetitive and menial tasks, and the use of software to replenish, forecast, and procure other processes. In IIoT, shop-floors comprise heterogeneous devices operating under different interfaces and different communication protocols. Additionally, the perceived data usually follow different data structures. Therefore, the fifth challenge is to establish interoperability among diverse platform-dependent devices and CPS to allow them to share information and interact with each other. To fuse the accessed data from various sources, data integration in terms of data cleaning or format conversion is also required.

During an industrial process, data is acquired from various  sources (such as physical and virtual assets) that are potentially owned by different organizations and is comprehensively aggregated to draw useful analysis. Such collection and collation of data from disparate sources need to address \textit{data management-related challenges}.
Since the data collected from the physical space is inputted to the virtual space,  the data management needs to determine the types of data to collect, how to organize them, and the data collection rate and resolution. What's more, supporting \enquote{track and trace solutions} throughout the industrial processes is required to promote transparency and fairness. Due to industry rivalry, commercial pressures dissuade companies from sharing sensitive information with their competitors. Therefore, managing the trade-off between transparency and privacy is another important challenge.

 \begin{figure*}[!ht]
\centerline{\includegraphics[width=6.5in]{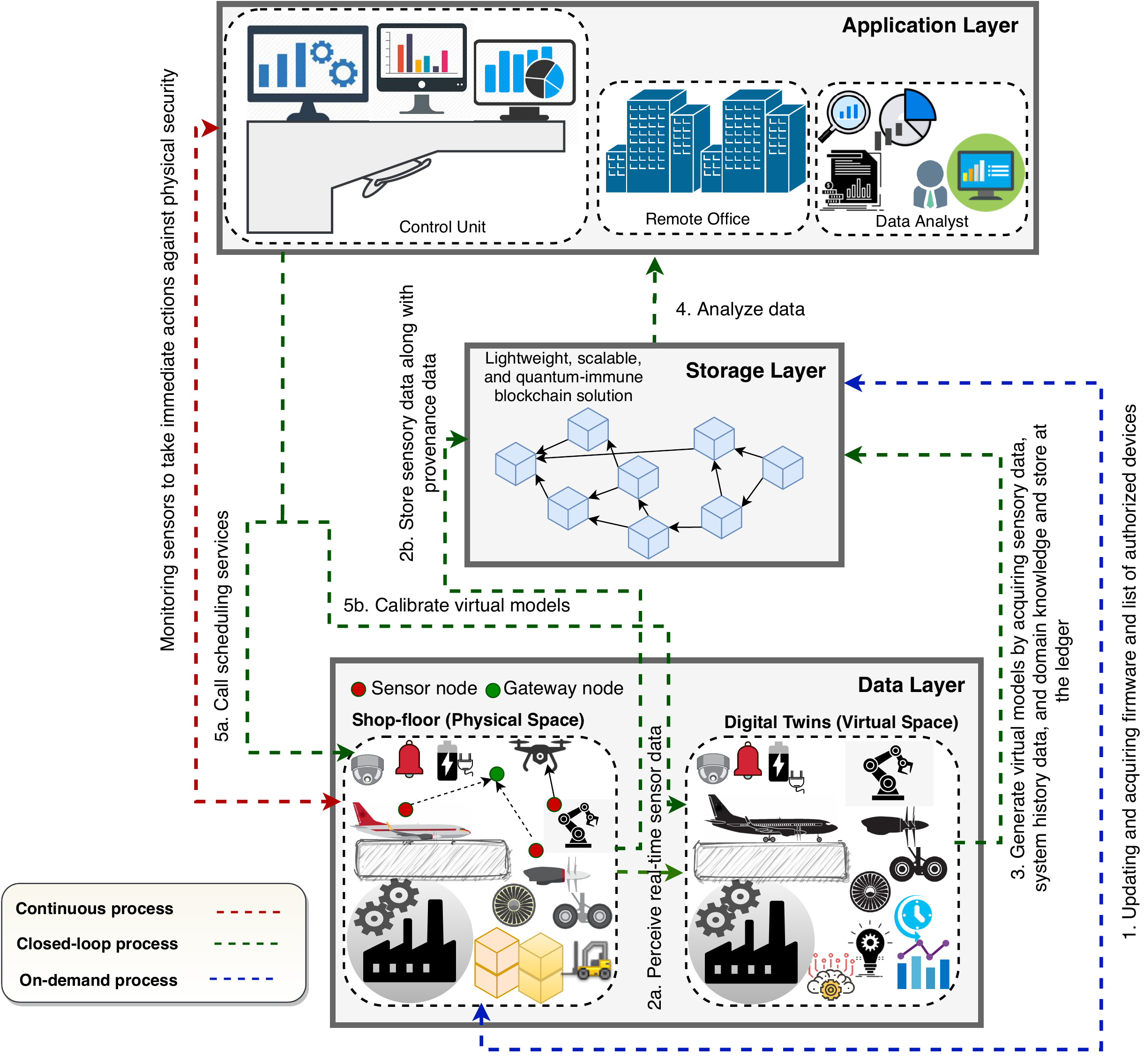}}
\caption{Blockchain with DTs: A framework illustrating monitoring, collection, storing, processing, and analysis of data from humans, machines, and sensing devices.}
\label{fig:IIoT}
\end{figure*}

Among \textit{data security-related challenges}, the most important challenge is to ensure data trustworthiness. Prior to the arrival of data at storage, data trustworthiness can be jeopardized by tampering or due to malfunctioning of data generating sources, and/or data modification by any of the intermediate rogue participating sources in IIoT. Such events may take place during supply chain processes or at manufacturing units where the malicious entities can be either external entities or authorized entities. The erroneous data (either maliciously or mistakenly) may lead to the insertion of false data into the storage thereby resulting in Garbage In Garbage Out (GIGO) problem. Other challenges include guaranteeing the confidentiality of sensitive sensor data or the company's trade secrets and enforcing distributed data accessibility and auditability based on ownership, roles, and access levels.

The \textit{performance-related challenges} cover issues such as maximizing the system throughput, ensuring deterministic and reliable data transfer and analysis in real-time for latency-sensitive scenarios, the optimal energy usage of resource-constrained devices, and minimizing the age of information to ensure the freshness of data.

\subsection{Non-technical Challenges}
The non-technical challenges cover formulating the best practices for identifying risk events (such as equipment deterioration, shrinkage, outage, natural disasters, etc.) and activating preventive and proactive plans such as aging management, fault diagnosis, or predictive maintenance accordingly to mitigate their effect. Additional challenges are related to cost management and retrofitting legacy systems.

While the aforementioned challenges are essential to be addressed;  we limit our scope to the exigent technical challenges pertaining to infrastructure, data management, and data security. In this regard, in the following section, we envision a blockchain-based framework for IIoT by leveraging DT. Furthermore, we integrate DT to address issues such as predictive maintenance, and anomaly detection. For illustration purposes, we consider an aircraft shop-floor use-case to map the significance of using blockchain and DTs. 

\begin{figure*}[ht!]
\centerline{\includegraphics[width=6.5in]{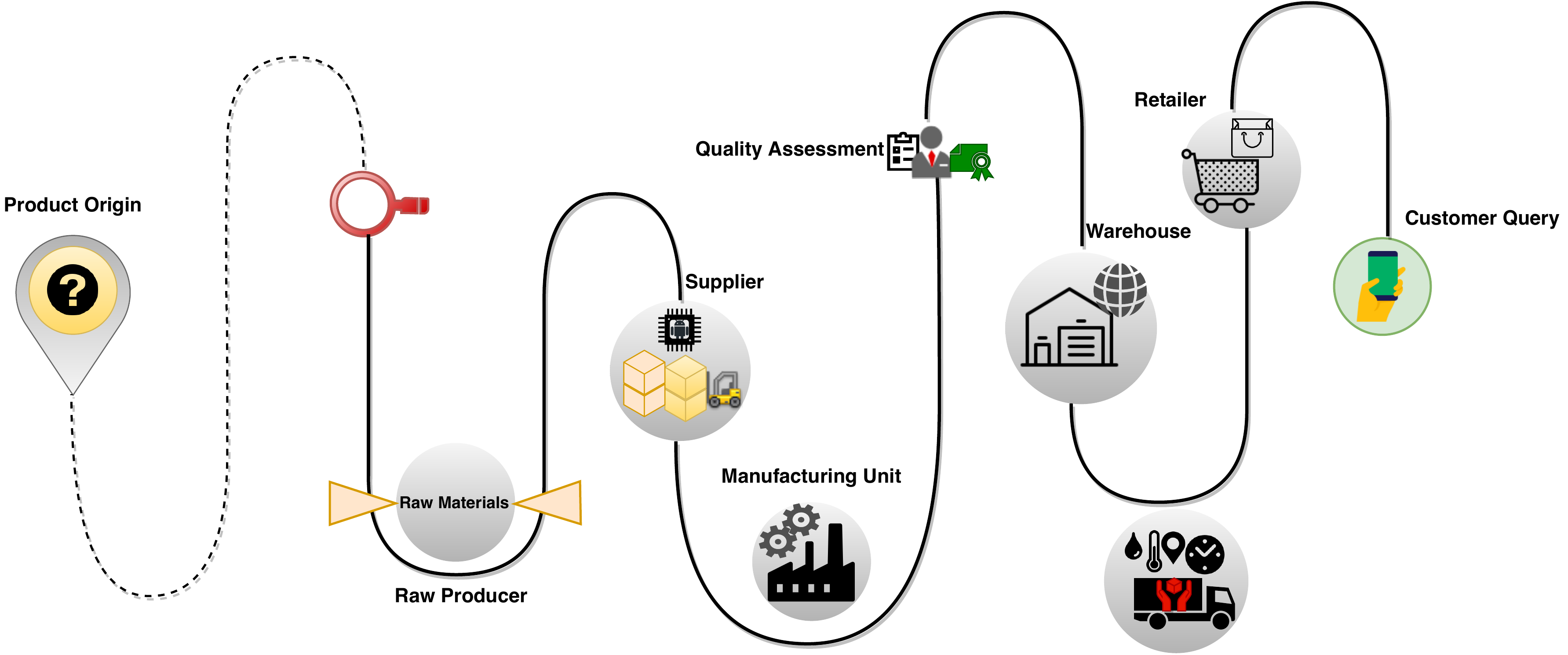}}
\caption{Product story as provenance data at each intermediary process in supply chain.}
\label{fig:prov}
\end{figure*}

\section{Integration of Blockchain and Digital Twin in IIoT}
\label{design_approach}
Before diving into further details, we discuss the main components of the proposed framework presented in Fig.~\ref{fig:IIoT}.
The participating entities (such as sensors, machines, and humans) initially register as authorized entities at the blockchain (step 1). Next, at the \textit{data layer}, assets monitor, collect, and process designated parameters in the physical space of the shop-floor (step 2a). The resource-constrained IoT devices monitor and collect data whereas the gateways and fog-enabled Unmanned Aerial Vehicles (UAVs) receive requests from sensors and process data. The collected data along with provenance is sent to the storage layer (step 2b). Provenance is a metadata that records a complete lineage of data and set of actions performed on data~\cite{suhail2020provenance}. For instance, provenance can answer when and where questions related to the product origin (as shown in fig.~\ref{fig:prov}). Based on the collected sensor data, domain knowledge, system history data, and process document, the twinned system generates models and store them at the storage layer (step 3). The \textit{application layer} keeps on analyzing data to detect any of the manufacturing disturbances (step 4). In case of any disturbance, the respective scheduling services (step 5a) in the physical space or model calibration services (step 5b) in the virtual space are called, completing the feedback loop. The \textit{storage layer} provides secure distributed data storage through a lightweight, scalable, and quantum-immune blockchain. The data can be accessed to and from blockchain, for instance, the application layer uses it for analysis and decision-making while DT uses it for generating and uploading virtual models. Additionally, to give priority to critical events, we deploy surveillance cameras, fire alarms, and power monitoring devices as physical security countermeasures. Since the repercussions of contingency events require immediate actions, therefore, we also enable the direct continuous monitoring of such events through sensory data logged at the control unit.

\begin{figure*}[ht!]
\centerline{\includegraphics[width=6.5in]{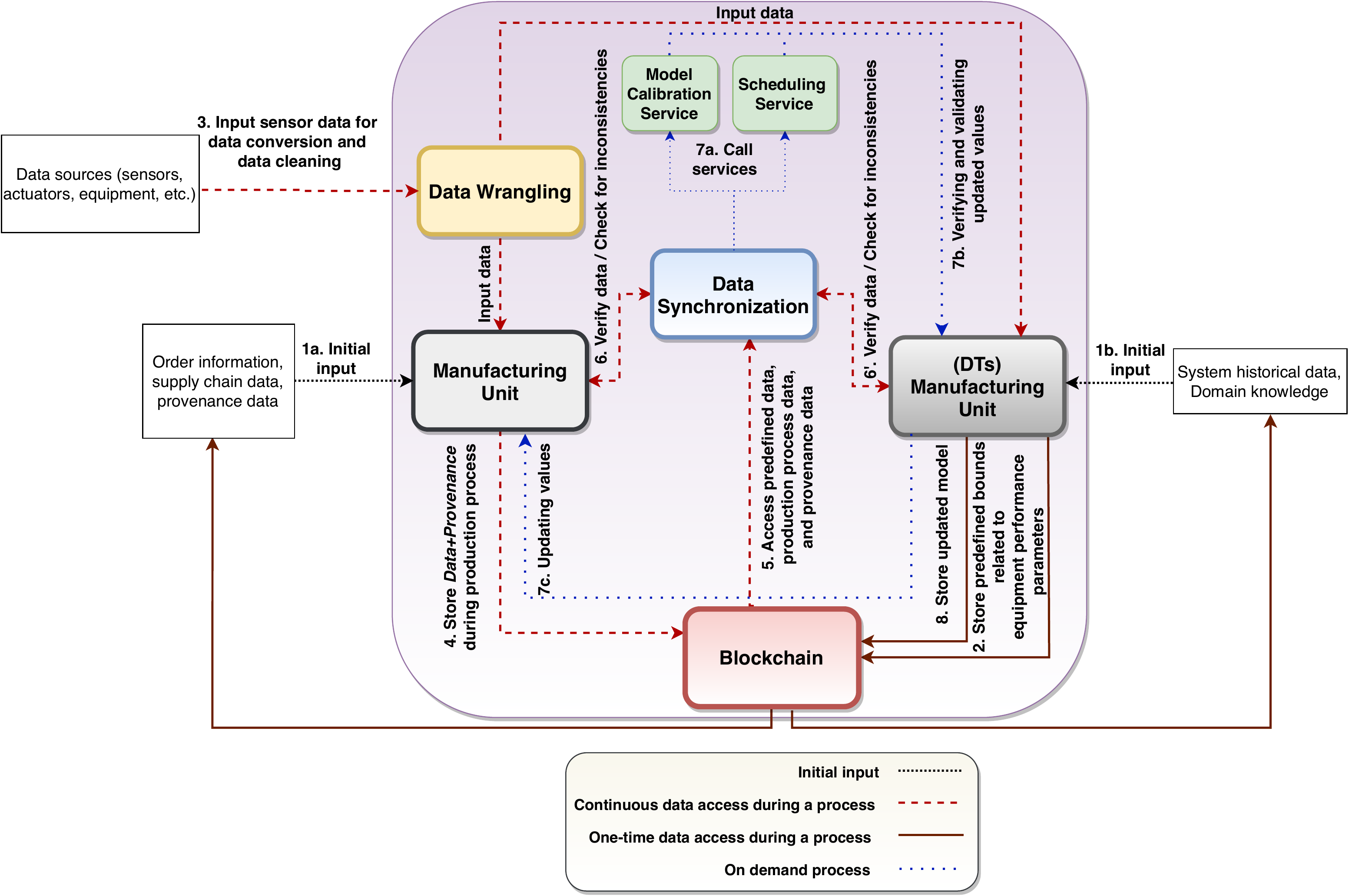}}
\caption{Data flow through the main components (manufacturing unit, digital twin, and ledger) and methods (data wrangling and data synchronization) of the proposed framework.}
\label{fig:flowdig_BC_DT}
\end{figure*}

To further elaborate the connection between the data layer and storage layer of the proposed framework, we refer to Fig.~\ref{fig:flowdig_BC_DT}. \textit{Before the production process}, the necessary details such as supply chain data (e.g., consignment information), order information (e.g., material stock, production quantity, estimated cost), simulation data (equipment historical data, prediction of equipment fault), etc. are already available in the storage system for usage (step 1a and 1b). Based on product lifecycle data, domain knowledge, and process document, a set of predefined values describing the acceptable ranges of system performance parameters are also stored in the storage system (step 2). We consider both steps 1 and 2 as one-time data access during the process. 

\textit{During the production process}, to enforce interoperability, we introduce a \textit{data wrangling} method responsible for cleaning invalid or missing data and converting different data formats into a unified format before inputting data to the twinned system (step 3).  As the process initiates, the underlying process data along with provenance is recorded on the storage system (step 4). Furthermore, we introduce a \textit{data synchronization} method (step 5) responsible for digital-physical mapping and checking for data inconsistencies. The reason for relying on such a process is to limit the frequent time-consuming access to the blockchain-based storage system, where we explicitly separate the data flow of real-time sensor data and the less dynamic production and provenance data.  
The data synchronization method performs a continuous mapping between the pre-defined equipment performance parameters, retrieved from the blockchain, and real-time sensor data, obtained from the manufacturing unit, to verify data consistency (step 6). The data synchronization method can access the updated process and provenance data directly from the ledger (other than the manufacturing unit) to eliminate the qualms of untrustworthy data. Note that, in case of data inconsistencies reported by the data synchronization method, corresponding scheduling services (in the physical space) or model calibration service (in the virtual space) carry out the necessary measures (step 7a) first at the virtual system (step 7b) and afterward regulate them on the physical system (step 7c). After resolving the issue, the updated model is stored in the blockchain (step 8).  

In the following, we discuss the challenges that can be addressed through the integration of DT and Blockchain in IIoT.

\subsection{Addressing Technical Issues in IIoT}
\subsubsection{Trustworthiness of data sources and data in transit}
Starting with the crucial requirements of data management and data security, the first step is to manifest the \emph{trustworthiness} of (i) data sources, and (ii) data in transit. Maliciously or mistakenly, data alteration may occur due to various reasons such as device tampering, data forging, or maliciously participating entity.

\begin{itemize}
    \item[a)] \emph{Proactive measures to circumvent device tampering}: To prevent device tampering, particularly for long-term functioning devices installed at critical infrastructures; sensors, gateways, and other types of digital equipment are bound to register as authorized devices through their public keys and digital certificates in the blockchain.
    Following such a strategy, the gateways collect data only from authorized devices by first confirming their authenticity from the ledger. Moreover, firmware updates or patch software of devices are also carried out through blockchain to prevent malware and other attacks related to the injection of malicious codes. For instance, storing the cryptographic hashes of device firmware helps to create a permanent record of device configuration and state. Upon verification that the software and settings have not been breached, the device is allowed to connect to other devices or services.
    
    \item[b)] \emph{Measures to diagnose data forgery}: To diagnose data forgery during data transition, we primarily rely on provenance data and twinned data. Firstly, data acquisition from the sources and data logging at the storage during each process can facilitate track and trace solutions through provenance data to identify the faulty entity in the underlying chained processes. Secondly, anomalies in the system can be detected based on digital-physical mapping in DTs. Usually, equipment performance parameters have desirable predefined bounds. A data anomaly occurs when the acquired data value(s) are out of those bounds. To identify such events, we categorize the sensor data as static and dynamic. The static data defines the range of sensor data (i.e., upper and lower bounds or pre-defined performance parameters of machines) while dynamic data represents the real-time sensor data. During the production process, a continuous comparison is carried out between the real-time sensor data generated from the physical device to the preset range in the twinned system for the identification of any accidental or deliberate abnormal behavior and data forging.
        
    \item[c)] \emph{Identification of rogue participating entity}: To identify a rogue participating entity, the following strategies can be adopted. Firstly, to meet the standards defined by the International Organization for Standardization (ISO), the regulatory endorsements can help to ascertain that industrial processes are abiding by ethical practices and environment-friendly operations. The exercising of such operations require the involvement of regulatory bodies (NGOs, governments, industry self-regulators) that conduct a periodic on-site inspection of the units to issue or revoke participation certificates. 
    Secondly, maintaining the reputation rating of the participating entities based on trust evaluation. In case of an authorized wrongdoer (usually during supply chain trade events), possible solutions involve the deployment of a reputation-based trading system~\cite{malik2019trustchain} or monetary punishment mechanism~\cite{ramachandran2017using} to reward honest entities while penalizing or revoking trader's participation.
  \end{itemize}

\subsubsection{Distributed, decentralized, and secure data storage}
Once we establish the reliability of data sources, the second step is to establish a distributed, decentralized, and secure data storage. In this regard, we leverage tamper-proof and immutable ledger of blockchain to procure distributed, decentralized, and secure data across multiple participating entities. In our proposed framework, data (either generated within the production facility or during the supply chain events), provenance data, DT data, and models, etc. flowing across multiple tiers or entities are stored and accessed through blockchain ledger. Through permissioned blockchain, we enable the coexistence of transparency and privacy to enforce confidential data flows and to cope with industrial espionage for competitive advantage.  Additionally, we enforce data accessibility by defining policies such that only authorized entities within the consortium are able to access and audit the data.

\subsubsection{Data traceability}
Blockchain solves the key issues, i.e., disparate data repositories and untrustworthy data dissemination; however, to reason about the current state of a data object such as \textit{why}, \textit{where}, and \textit{how}
entails a complete lineage of processes chain. In our proposed framework, we use provenance data to identify the cause of deviations between the physical-digital space in addition to data assessment, such as data debugging, auditing, reconciliation, performance tuning, etc. 

\subsubsection{Scalability}
The scalability snag associated with the mainstream sequential chain-structured blockchains, such as Bitcoin, Ethereum, and Hyperledger is expected to be settled by scalable solutions, such as DAG-structured blockchains (IOTA~\cite{popov2017iota}, Byteball~\cite{churyumov2016byteball}, NANO~\cite{lemahieu2018nano}) and chain-structured blockchains (LSB~\cite{LSB}, Tree-Chain~\cite{Tree-Chain}, PoET~\cite{PoET}). 
Therefore, considering the futuristic surge based on the growing number of actors and activities in the network, we adopt DAG-based blockchains in our proposed design.

\subsubsection{Quantum immunity}
The blockchain fulfills the principles of \emph{Confidentiality, Integrity, and Availability (CIA) triad} through hash functions, public-key cryptosystems and data replication respectively. However, the security of de-facto cryptographic primitives is at risk of being broken by quantum computers~\cite{cheng2017securing}. For example, quantum computers put blockchain security at risk by enabling wrongdoers to forge digital signatures and sabotage transactions~\cite{fedorov2018quantum}. Therefore, to prepare for the quantum apocalypse, blockchain schemes that inherently support post-quantum cryptography, such as IOTA are adopted in our proposed design~\cite{suhail2020role}.

\subsubsection{Optimal energy consumption of resource-constrained devices}
For highly energy-constrained IoT devices, such as battery-powered, performing computationally expensive tasks is not practically possible without hardware-accelerated cryptography. In the proposed framework, we assign the usual tasks, such as monitoring and collecting data to the resource-constrained IoT devices in order to conserve sensor energy. The gateways and fog-enabled UAVs perform power-intensive tasks, such as receiving requests from sensors, processing data, or broadcasting transactions in the network.

\subsubsection{Interoperability}
In IIoT, interoperability is another challenging issue that arises due to the integration of multiple internal and external (sub) systems. In our case, we primarily focus on syntactic interoperability that arises due to different formats and the data structure used in any exchanged information or service. Before the gathered data is fed into the twinned system, data conversion and data cleaning through the data wrangling method transform the data into a unified form (as shown in Fig.~\ref{fig:flowdig_BC_DT}).

\section{Research Challenges and Open Issues}
In the following, we identify and discuss pressing issues towards the successful implementation of DTs and blockchains in IIoT.

\subsection{Accurate representation of digital footprints}
One of the critical challenges for a virtual model representing the real-world situation is \textit{high-fidelity}. In production environments, the ideal DT (for example, for a robotic arm) should mirror an exact image of all the properties and functions of the physical component, synchronized in real-time throughout its life. However, considering the variability, fuzziness, and uncertainty of physical space how to accurately model DTs is still an open issue.

\subsection{Big data challenges} 
Data challenges encompass two main factors: (i) What types of data to collect and how frequently?, and (ii) where to store data?  
The former issue entails a trade-off between excessive data and limited data, i.e., too few, inaccurate predictions; too many, mired down in details or pose decision paralysis due to information overload. Furthermore, factors such as the frequency at which data samples are collected and the duration for which they are stored need to be determined as they are linked with high storage cost and long-term data required for modeling respectively~\cite{kusiak2017smart}. 
The latter issue corresponds to storing \textit{big data} on blockchain ledger that requires careful consideration of factors such as high cost associated with high volumes of rapid measurements, block size, latency, etc. Alternative solutions are off-chain storage (such as P2P file systems) or storing data hash only to reduce storage costs.
Removable blockchain designs, as in~\cite{dorri2019mof}, allow removal of transactions from blockchains without breaking its consistency.  Even though blockchains have significantly solved the data management issues, the problems associated with the storage of big data keeping in view the volume and velocity of big data are still open issues.

\section{Conclusion}
In IIoT, ensuring the trustworthiness of data is a challenging task that requires applying essential measures throughout the process of data generation and propagation. In this article, we focus on such challenges.
Starting with our first objective to ensure the reliability of data sources, we employ track and trace solutions through provenance data. Then we enforce the use of a secure distributed blockchain ledger. Finally, driven by data collected and accessed from reliable data sources and data storage respectively, we make use of DTs models that mirror every facet of a product and thus help to achieve predictive maintenance. Furthermore, we identify future research challenges with an open-ended discussion in the adoption of blockchain and DTs by the IIoT.

\bibliographystyle{IEEEtran}

\begin{thebibliography}{10}
\providecommand{\url}[1]{#1}
\csname url@samestyle\endcsname
\providecommand{\newblock}{\relax}
\providecommand{\bibinfo}[2]{#2}
\providecommand{\BIBentrySTDinterwordspacing}{\spaceskip=0pt\relax}
\providecommand{\BIBentryALTinterwordstretchfactor}{4}
\providecommand{\BIBentryALTinterwordspacing}{\spaceskip=\fontdimen2\font plus
\BIBentryALTinterwordstretchfactor\fontdimen3\font minus
  \fontdimen4\font\relax}
\providecommand{\BIBforeignlanguage}[2]{{%
\expandafter\ifx\csname l@#1\endcsname\relax
\typeout{** WARNING: IEEEtran.bst: No hyphenation pattern has been}%
\typeout{** loaded for the language `#1'. Using the pattern for}%
\typeout{** the default language instead.}%
\else
\language=\csname l@#1\endcsname
\fi
#2}}
\providecommand{\BIBdecl}{\relax}
\BIBdecl

\bibitem{suhail2019orchestrating}
S.~Suhail, R.~Hussain, C.~S. Hong, and A.~Khan, ``Orchestrating product
  provenance story: When {IOTA} ecosystem meets the electronics supply chain
  space,'' 2019, {a}rXiv: 1902.04314. [Online]. {A}vailable at:
  \url{https://arxiv.org/abs/1902.04314}.

\bibitem{yaqoob2020blockchain}
I.~{Yaqoob}, K.~{Salah}, M.~{Uddin}, R.~{Jayaraman}, M.~{Omar}, and M.~{Imran},
  ``Blockchain for digital twins: Recent advances and future research
  challenges,'' \emph{IEEE Network}, vol.~34, no.~5, pp. 290--298, 2020.

\bibitem{tao2018digital}
F.~Tao, J.~Cheng, Q.~Qi, M.~Zhang, H.~Zhang, and F.~Sui, ``Digital twin-driven
  product design, manufacturing and service with big data,'' \emph{The
  International Journal of Advanced Manufacturing Technology}, vol.~94, no.
  9-12, pp. 3563--3576, 2018.

\bibitem{tao2017digital}
F.~Tao and M.~Zhang, ``Digital twin shop-floor: a new shop-floor paradigm
  towards smart manufacturing,'' \emph{{IEEE} Access}, vol.~5, pp.
  20\,418--20\,427, 2017.

\bibitem{deloitte}
Deloitte, ``Continuous interconnected supply chain,'' available at:
  \url{https://www2.deloitte.com/
  content/dam/Deloitte/lu/Documents/technology/lublockchain-internet-things-supply-chain-traceability.pdf.}

\bibitem{suhail2020provenance}
S.~Suhail, R.~Hussain, M.~Abdellatif, S.~R. Pandey, A.~Khan, and C.~S. Hong,
  ``Provenance-enabled packet path tracing in the {RPL}-based internet of
  things,'' \emph{Computer Networks}, p. 107189, 2020.

\bibitem{malik2019trustchain}
S.~Malik, V.~Dedeoglu, S.~S. Kanhere, and R.~Jurdak, ``Trustchain: Trust
  management in blockchain and {IoT} supported supply chains,'' \emph{arXiv
  preprint arXiv:1906.01831}, 2019.

\bibitem{ramachandran2017using}
A.~Ramachandran, D.~Kantarcioglu \emph{et~al.}, ``Using blockchain and smart
  contracts for secure data provenance management,'' 2017, {a}rXiv: 1709.10000.
  [Online]. {A}vailable at: \url{https://arxiv.org/abs/1709.10000}.

\bibitem{popov2017iota}
S.~Popov, ``The tangle. {W}hite paper.'' \emph{Available at:
  \url{https://iota.org/IOTA_Whitepaper. pdf}, 2016}.

\bibitem{churyumov2016byteball}
A.~Churyumov, ``Byteball: A decentralized system for storage and transfer of
  value,'' {B}yteball.org, Moscow, Russia, 2017. [Online]. {A}vailable at:
  \url{https://obyte.org/Byteball.pdf}.

\bibitem{lemahieu2018nano}
C.~LeMahieu, ``Nano: A feeless distributed cryptocurrency network,'' 2018,
  [Online]. {A}vailable at: \url{https://nano.org/en/whitepaper}.

\bibitem{LSB}
A.~Dorri, S.~S. Kanhere, R.~Jurdak, and P.~Gauravaram, ``{LSB}: A lightweight
  scalable blockchain for {IoT} security and anonymity,'' \emph{Journal of
  Parallel and Distributed Computing}, vol. 134, pp. 180--197, 2019.

\bibitem{Tree-Chain}
S.~{Malik}, V.~{Dedeoglu}, S.~S. {Kanhere}, and R.~{Jurdak}, ``Trustchain:
  Trust management in blockchain and {IoT} supported supply chains,'' in
  \emph{2019 IEEE International Conference on Blockchain (Blockchain)}, 2019,
  pp. 184--193.

\bibitem{PoET}
``Hyperledger sawtooth,'' available at:
  \url{https://www.hyperledger.org/use/sawtooth}.

\bibitem{cheng2017securing}
C.~Cheng, R.~Lu, A.~Petzoldt, and T.~Takagi, ``Securing the internet of things
  in a quantum world,'' \emph{IEEE Communications Magazine}, vol.~55, no.~2,
  pp. 116--120, 2017.

\bibitem{fedorov2018quantum}
A.~K. Fedorov, E.~O. Kiktenko, and A.~I. Lvovsky, ``Quantum computers put
  blockchain security at risk,'' \emph{Nature}, vol. 563, pp. 465--467, 2018.

\bibitem{suhail2020role}
S.~{Suhail}, R.~{Hussain}, A.~{Khan}, and C.~S. {Hong}, ``On the role of
  hash-based signatures in quantum-safe internet of things: Current solutions
  and future directions,'' \emph{IEEE Internet of Things Journal}, pp. 1--1,
  2020.

\bibitem{kusiak2017smart}
A.~Kusiak, ``Smart manufacturing must embrace big data,'' \emph{Nature}, vol.
  544, no. 7648, pp. 23--25, 2017.

\bibitem{dorri2019mof}
A.~Dorri, S.~S. Kanhere, and R.~Jurdak, ``Mof-bc: A memory optimized and
  flexible blockchain for large scale networks,'' \emph{Future Generation
  Computer Systems}, vol.~92, pp. 357--373, 2019.

\end{thebibliography}

\section*{Biographies}

\vskip -2.5\baselineskip plus -1fil

\begin{IEEEbiography}{Sabah Suhail} is a Ph.D. scholar at Kyung Hee University, Korea. Her research interests include blockchain, provenance, security and privacy issues in the Internet of Things.
\end{IEEEbiography}

\vskip -2.5\baselineskip plus -1fil
\begin{IEEEbiography}{Rasheed Hussain} is Associate Professor, Director of Networks and Blockchain Lab, and Head of MS program in Security and Network Engineering at Innopolis University, Russia. His research interests include information, cyber, and network security, privacy, vehicular networks, vehicular clouds, vehicular social networking, applied cryptography, Internet of Things, content-centric networking, cloud computing, blockchain, API security, and machine learning in cybersecurity.
\end{IEEEbiography}

\vskip -2.5\baselineskip plus -1fil
\begin{IEEEbiography}{Raja Jurdak} is Professor and Chair at Queensland University of Technology, and Director of the Trusted Networks Lab.  His research interests include trust, mobility and energy-efficiency in networks. Prof. Jurdak has published over 170 peer-reviewed publications, including two authored books most recently on blockchain in cyberphysical systems in 2020. He serves on the editorial board of Ad Hoc Networks, and on the organising and technical program committees of top international conferences, including Percom, ICBC, IPSN, WoWMoM, and ICDCS. 
\end{IEEEbiography}

\vskip -2.5\baselineskip plus -1fil
\begin{IEEEbiography}{Choong Seon Hong} is Professor at Kyung Hee University, Korea. His research interests include future Internet, ad hoc networks, network management, and network security. 
\end{IEEEbiography}

\end{document}